# Understanding and Evaluating Trust in Generative AI and Large Language Models for Spreadsheets

Dr Simon Thorne, Cardiff School of Technologies, Cardiff Metropolitan University

sthorne@cardiffmet.ac.uk

**ABSTRACT**

Generative AI and Large Language Models (LLMs) hold promise for automating spreadsheet formula creation. However, due to hallucinations, bias and variable user skill, outputs obtained from generative AI cannot be assumed to be accurate or trustworthy. To address these challenges, a trustworthiness framework is proposed based on evaluating the transparency and dependability of the formula. The transparency of the formula is explored through explainability (understanding the formula's reasoning) and visibility (inspecting the underlying algorithms). The dependability of the generated formula is evaluated in terms of reliability (consistency and accuracy) and ethical considerations (bias and fairness). The paper also examines the drivers to these metrics in the form of hallucinations, training data bias and poorly constructed prompts. Finally, examples of mistrust in technology are considered and the consequences explored.

**1.0 INTRODUCTION**

This paper explores the concept of trust in generative AI and Large Language Models (LLMs) for spreadsheet formulas generation and proposes a series of dimensions that could be used to evaluate trust in AI generated spreadsheet artefacts.

The research questions used to guide this work are as follows:

1. What significant differences and similarities are there between trust dimensions for automatic machines and generative AI and LLMs?
2. What sources of error and threats to trust exist in generative AI and LLMs and how might these issues be mitigated?
3. Can dimensions of trust in automation be adapted to evaluate spreadsheet formulas generated by AI and LLMs?

The aim of this work is to provide some starting point for the development of a trustworthiness framework that could be applied to generative AI and LLM produced spreadsheet formulas considering the specific challenges of spreadsheets.

**1.1 Understanding the role of generative AI and LLMs in spreadsheets**

Generative AI has recently advanced to a point where it can be leveraged for a variety of different tasks including computer programming. Most generative AI systems can produce computer code in a variety of languages, including spreadsheets. Generative AI is now part of many mainstream software applications in the form of Copilot for Microsoft products and countless programming assistant tools based on GPT and other LLMs. The impact of such tools in spreadsheets remains to be fully seen but there is a steady increase in research that leverage LLMs for spreadsheets. Some research offers alternative generative strategies for creating entire spreadsheet applications rather than just formulas (Li, et al. 2024), some offers special cases for spreadsheets such as Copilot for Finance (Microsoft 2024) or prompt and auditing support in the form co-audit for Excel (Gordon, et al. 2023).

**1.2 Similarities between spreadsheets and generative AI**

The arrival, impact and implications of generative AI are analogous to the introduction of the spreadsheet. Spreadsheets were and are still prevalent for a variety of reasons including affordability of access, flexibility and their universal nature. Spreadsheets are utterly ubiquitous and have become the "dark matter and energy" (Panko 2013) of corporate IT. The amount of data processed in spreadsheets eclipses that of corporate IT systems and this trend is only likely to continue.

Generative AI is prevalent for many of the same reasons as spreadsheets but provides an even greater reach; for instance being able to generate programming code in a variety of languages. Generative AI will become entrenched in many industries with new applications and new domain specific generative AI tools. But like spreadsheets, they are also a potential source of error and require careful consideration to ensure their reliability and accuracy in order to objectively evaluate the trust that should be placed in them.

**2.0 TRUST IN AUTOMATION**

Trust as a concept is a critical dimension to human decision-making. To trust an individual is to grant the assumption of reliability and predictability to work toward a common goal. Trust is granted over time and is based on objective observations that can be made about the intent, reliability and predictability of a collaborator.

Trust in automation - the trust given to an automatic machine - has similar dimensions to other kinds of trust that humans develop. Muir (Muir and Moray 1996, B. Muir 1994) defines trust in automation as "*an attitude that an agent will help achieve an individual's goals in a situation characterised by uncertainty and vulnerability*"

Muir presents a model for evaluating trust based on automatic factory machines, this work has been adapted to suit computers and to model human computer interaction but given the novelty of generative AI, it feels appropriate to go back to first principles. There have been many models of trust in automation developed by researchers, a good overview of those models and the metrics used are given by (Kohn, et al. 2021). Most models of trust use a number of different dimensions to evaluate dynamic trust. These dimensions include but are not limited to: Reliability, Transparency, User Training/Experience, Predictability and User Feedback.

For the purposes of evaluating trust in generative AI for spreadsheet formula generation, a set of novel trust dimensions and metrics are proposed, greatly influenced by a number of researchers (Muir and Moray 1996, B. Muir 1994, Lee and See 2024) and attempts to take in the nuance of generative AI processes which differ from the contexts described in the research used as inspiration.

**2.1 Transparency and trustworthiness framework for generative AI spreadsheet formula**

Given the unique qualities, attributes and mechanisms at play in generative AI, the following metrics are proposed for evaluating trust in generative AI spreadsheet formula. These metrics are designed to give users some objective metrics for dimensions of trust tailored to generative AI formula production. All of these metrics together provide a "Transparency and Trustworthiness Framework".

### 2.1.1 Transparency:

**Explainability:** Generative AI should provide clear explanations for the formulas it generates. Users need to understand the logic and reasoning behind each recommendation to trust its outputs. Explainability mechanisms, such as providing step-by-step explanations or highlighting key factors influencing formula generation, enhance transparency and can be defined when engineering the prompt. Users could also adopt explainability benchmarks for objective evaluation (White, et al. 2023, Lee and See 2024, Ying, et al. 2024).

**Visibility**: Generative AI's algorithms should be transparent to users, allowing them to inspect and understand its inner workings. Visibility differs from explainability since it focuses on the underlying computational processes and not the generative AI reasoning presented to the user. Providing visibility into the AI's processes, such as disclosing the model architecture, training data, and parameters, fosters trust by enabling users to scrutinise its behaviour and performance. Visibility can be engineered in the prompt to provide some of these features but due to the manner in which generative AI is trained, some of this information is "unknowable" due to deep learning neural networks being "black boxes", there are some approaches to mitigating these issues which could provide causal explanations (Bhattacharjee, et al. 2023).

### 2.1.2 Dependability:

**Reliability**: Generative AI should consistently produce accurate and reliable formulas that meet users' expectations. Rigorous testing, validation, and quality assurance processes ensure formula generated by AI are accurate and agree with real world observations. One such approach that could help users determine the reliability of the algorithm is the development of a series of benchmark tests which generative AI is regularly tested against. These benchmarks could be themed around known issues in LLMs, for instance input uncertainty in the prompt, prompts that deal with negation, or prompts that require deduction or inference. For spreadsheets, a good starting point for these benchmarks is explored in Thorne (2023) and O'Beirne (2023) and beyond spreadsheets there are numerous approaches to benchmarking generative AI and LLMs (Ye, et al. 2023).

**Ethical Considerations:** Generative AI should adhere to ethical principles and guidelines, respecting users' privacy, autonomy, and well-being. Ethical design practices to ensure fairness, accountability, and transparency in algorithmic decision-making can enhance trustworthiness by addressing users' concerns about potential biases or discriminatory outcomes. Privacy preserving techniques such as differential privacy could be employed to protect user data. Bias and fairness issues can be evaluated through different tools such as IBM's AI Fairness tool which detects and mitigates against algorithmic bias (Bellamy, et al. 2019). Users could also establish a regular bias audit and produce documentation to track instances of detected bias in generated output.

### 2.1.3 User-Centric Design:

**User Feedback Mechanisms:** Generative AI should incorporate feedback loops that enable users to provide input, review outputs, and suggest improvements. User feedback mechanisms, such as rating systems or interactive interfaces for refining formulas, engage users in the formula generation process and empower them to contribute to the improvement of the tool. It would also be useful for LLMs to indicate the level of probability in their responses, this is a natural byproduct of the technical process by which LLMs compute their responses. For each formula that is produced, a confidence figure could also be obtained to make it clearer where the output is generated with a low probability of it being correct. This can be incorporated into the prompt used to generate the output and would be an invaluable indicator of potential quality or hallucination issues.

## 2.2 Conclusions on Transparency and Trustworthiness Framework

The above discussion provides some overview of how different approaches to exploring dimensions of generative AI can be used to try and objectively determine the trustworthiness of a particular generative AI artefact. The suggested areas are pragmatic although a wraparound process would be required to effectively implement such a strategy. The proposals in this section are simply a starting point; a means of trying to map out some of the important issues in determining the level of trust that should be placed in generative AI artefacts. These metrics of trust are designed to combat the sources of error that make generative AI inherently difficult to trust. These sources of error, threats to trust and consequences of mistrust are outlined in the next section.

## 2.3 Sources of error and threats to trust

Sources of error and threats to trust arise from the user or from the technical process by which generative AI and LLMs work. Hallucinations and bias arise through the process of training generative AI and LLMs. Magical thinking, reification and deficiencies in prompt engineering arise in users and relate to how generative AI is perceived and the knowledge that the user possesses.

### 2.3.1 Hallucinations

Hallucinations are defined as:

> *"LLM hallucinations are the events in which ML models… produce outputs that are coherent and grammatically correct but factually incorrect or nonsensical. "Hallucinations" in this context means the generation of false or misleading information. These hallucinations can occur due to various factors, such as limitations in training data, biases in the model, or the inherent complexity of language."* (IBM 2024)

Hallucinations represent the biggest threat to the usefulness and utility of generative AI because they can be subtle and are generally presented in an authoritative tone regardless of accuracy. The user must be able to validate and verify in the output in the knowledge domain of the application, meaning that they must have sufficient knowledge to know if the output is sensible. Otherwise, the user is left without an objective means to be sure that the output is not a hallucination.

Hallucinations seem to be triggered by certain conditions present in the prompt. Uncertainty, the need for deduction or inference, negation and some mathematical operations all appear to trigger hallucinations in LLMs (Chen, et al. 2023, Plevris, Papazafeiropoulos and Jiménez Rios 2023, Wan, et al. 2024, Shakarian, et al. 2023, Davis 2023, Frieder, et al. 2023, O'Beirne 2023). The result is an output that is inaccurate, flawed proofs or reasoning or an artefact that bears little relevance to the prompt provided.

### 2.3.2 Magical thinking and Reification

Magical thinking is defined as *"Making causal connections or correlations between two events not based on logic or evidence, but primarily based on superstition"* (Sternberg, Roediger and Halpem 2020) in more broad terms magical thinking is a *"blurring between cause and effect which can result in giving agency to something or someone which is not deserving"* (Krieshok 2020).

In the case of generative AI, this may manifest itself in non-factual beliefs about what generative AI is capable of, for instance a sweeping generalisation like the belief that generative AI is never incorrect. Alternatively, it may be more subtle, for instance the belief that through the process of training the model on human writing, that generative AI now possess that "knowledge" and as such extend human properties to generative AI such as wisdom. Of course, generative AI and LLMs possess no knowledge, they can only provide a probable answer based on the input.

Reification is similar to magical thinking. It is where an abstract idea, like a computer application or data model is transformed by trust into a "real" object that becomes the unquestionable truth. Users of those systems perceive the reified system as the embodiment of that idea and as such is bestowed with properties such as *believability, correctness, appropriateness, concreteness, integrity, tangibility, objectivity and authority* (Croll 2017). Reification can occur through a legitimate and correct perception of reliability, through a distorted and incorrect view of apparent reliability, or it can be asserted without the need for any evidence. In conjunction with magical thinking, there is a significant chance that generative AI or analytical products of generative AI may become reified simply because of the means by which they were generated.

### 2.3.3 Bias

Bias is another significant issue in LLMs due to the process by which the model is trained. LLMs are trained on a "vast" corpus of human writing. The exact makeup of this "vast" training corpus is a "proprietary secret" (according to ChatGPT4) with OpenAI refusing to provide these details. ChatGPT3.5 has 175 Billion parameters and was trained on "hundreds of billions" of words (Brown, et al. 2020), presumably ChatGPT4 has more tuneable parameters and is trained on more words but this is a closely guarded secret. Inevitably, this human writing contains different kinds of conscious and subconscious bias such a misogyny or racism (Fang, et al. 2024, Lucy and Bamman 2021, Zack, et al. 2024). Overt examples of bias are removed as part of the review of the training process, but subconscious issues are by their nature subtle and hard to identify, hence there is a chance that this bias can creep into generative AI and LLM outputs including code generation (Haung, et al. 2024).

### 2.3.4 Deficiencies in prompt engineering

Prompt engineering is an emerging discipline that aims to guide generative AI and LLMs to *better* outputs through offering greater specificity in the prompt which in turn should reduce hallucinations and provide artefacts that are closer to the intention of the user. The skill of a user in prompt engineering depends on their knowledge of prompt engineering techniques, their domain knowledge, their ability to plan ahead of generating the output and their ability to validate and verify the output. In a similar way to spreadsheet modelling or software coding, if the programmer has an unclear idea of what the requirements are, or they only have a partial view of the requirements, the resulting software is likely to be unsatisfactory.

There are many prompt libraries with existing prompts that can be leveraged in novel situations. These libraries act as templates which can be adapted to the scenario. Some prompt engineering libraries are domain specific for instance in medicine (Ahmed, et al. 2024, Heston and Khun 2023, Wang, et al. 2024) or general purpose and non-domain specific (Lo, 2023, Marvin, et al. 2024, Lo, 2024, Stroblet, et al. 2023). Knowledge of these methods is critical to avoiding hallucinations and writing *better* prompts.

Planning could have a significant impact on the quality of the prompt through avoiding omissions. Omission errors in constructing the prompt are likely to increase ambiguity and uncertainty which drives hallucinations or simply insufficient outputs. Performing some basic planning such as producing a mind map or a more technical approach like data flow diagrams could reduce the likelihood of omissions in the prompt. These methods could achieve this through forcing the user to think about information flow and the input and output of the problem they are trying to generate code for. This should reduce the likelihood of basic omissions being made which in turn should improve the quality of the output obtained from generative AI and LLMs. This would appear to be particularly true where the user has deficiencies in domain knowledge but it would require users to research and explore the domain sufficiently.

Further, agile methods like Test Driven Development (TDD) could be adapted to help drive generative AI and LLM output based on specific input/output test cases (Mathews and Nagappan 2024). Little research exists at present on the effectiveness of planning methods on prompt quality, yet it would it seems to warrant significant attention.

User domain knowledge is another important factor to consider as those operating generative AI and LLMs in familiar domains will perform better than those operating in unfamiliar domains. Domain knowledge is critical in constructing specific prompts and avoiding ambiguity which may drive hallucinations. It is also critical in validating and verifying outputs of generative AI and LLMs, without sufficient domain knowledge it may be impossible to verify the output. Little research exists on the impact of domain knowledge on generative AI and LLM prompt quality, but it would seem to be an important consideration.

### 2.4 Mistrust

Mistrust describes a scenario where trust is incorrectly placed in a system which eventually fails, this trust elevates the system to position of authority with the default assumption being made about the system is that it is accurate and worthy of this status. Mistrust arises through failure to fully evaluate all of the dimensions of trust which in turn mask lurking problems, for instance failing to stress test or verify the operation of the formula or wider system. Bugs and errors can also remain unnoticed until the right conditions are met to trigger these failures. A failure to plan, build and test in a rigorous way will increase the risk that the trust placed in a system is misplaced. Unfortunately, there are an abundance of examples where catastrophic failure reveals the trust placed in a system was misplaced.

#### 2.4.1 Reinhart and Rogoff

In the wake of the financial crisis in 2008, two Harvard professors published a paper "Growth in a time of debt" (Reinhart and Rogoff 2010) that performed an analysis on GDP to debt ratios for various countries to determine the optimal debt to GDP ratios. The paper asserted that once a country exceeded 90% debt to GDP ratio, the only outcome was negative growth. This conclusion was used by governments around the world to justify austerity politics, George Osbourne, the Chancellor in the UK at the time, even made direct reference to the 90% limit in parliament. However, the analysis was flawed because several rows of data in the analysis spreadsheet had been accidentally omitted from the calculation. If these rows had been included, the only conclusion that could be reached is that there is no hard limit between debt and GDP ratios and hence the justification for austerity had no basis in fact.

There was never any correction of austerity politics or even acknowledgement that the evidence used to justify this course of action was false. Research shows that between 2012 and 2019, there were 335K excess deaths in the UK which have been attributed to austerity politics and a general increase in mortality across high income nations (Walsh, et al. 2022, McCartney, et al. 2019).

### 2.4.2 UK Test and Trace

In 2020 during the height of the global COVID19 pandemic, it was reported that the UK £42B test and trace system had 'lost' 16,000 positive test cases. The test and trace system batch processed positive test cases from various authorities in the UK. This data was collected locally and then combined to produce overall statistics reported by the UK government and the Scientific Advisory Group for Emergencies (SAGE). It is unclear precisely what happened but either data was imported into an older format of Excel (.xls) most likely from a comma separated value (CSV) file or somewhere in the pipeline of data flow, there was an older version of Excel using the .xls format. This format could only represent 65,000 rows of data. When a file larger than this is encountered, the bottom part of file is simply chopped off, apparently without warning or raising an obvious error message, accounting for the 16,000 lost test cases.

Those lost test cases meant that individuals with a positive diagnosis of COVID-19 were unaware of their condition and as such were not taking relevant precautions to protect others and those who encountered these individuals would not have been alerted to their exposure either. Failing to trace these 16,000 COVID positive cases may have caused an additional 1500 deaths (Fetzer and Graeber 2021).

### 2.4.3 The Post Office Horizon Computer Scandal

The British Post Office Horizon scandal involved the wrongful prosecution of over 900 sub-postmasters from 1999 to 2015, due to errors in the Horizon accounting software developed by Fujitsu. Despite knowing about the software bugs, the Post Office insisted on Horizon's reliability, leading to convictions for theft and fraud, financial ruin, and severe personal consequences, including suicides. The scandal came to light in 2009, prompting legal action and a 2019 High Court ruling that the contracts were unfair and the software faulty. Subsequent investigations and court rulings led to the quashing of 100 convictions by February 2024, with compensation costs expected to exceed £1 billion. These convictions expose frailties in the legal system, specifically the presumption of reliability in electronic evidence, here the trust placed in the reliability of computer evidence led to this scandal although the case is complex with many contributing factors (Mason and Seng 2021, MasonA 2023, MasonB 2023, Christie 2020). In May of 2024, the UK government passed into law, the Post Office (Horizon System) Offences Act which quashed all of the convictions of subpostmasters (Post Office 2024).

### 2.4.4 The consequences of mistrust

The examples cited in the previous section of misplaced trust are a few of many. These cases are picked to illustrate that even when the system is safety critical, such as the test and trace system or the implications are far reaching, such as in the case of Reinhart and Rogoff and Horizon, there no guarantees that the importance of the task ensures that proper process and precautions are taken to ensure accuracy and trustworthiness even if the consequences are severe.

In the case of Reinhart and Rogoff, the analysis provided was trusted at many levels. The source of the assertion and analysis would appear to be trustworthy as it was written by two Harvard economics Professors. The analysis was published in a leading journal following peer review and the contents of the paper were trusted by decision makers who used the evidence to justify economic strategies affecting hundreds of millions worldwide.

In the case of COVID19, the software and data pipelines were trusted to be reliable and predictable since surveillance of the progression of the virus was a critical part of the strategy to manage the COVID19 outbreak in the UK.

In the Horizon scandal, trust was extended to the reliability of the electronic evidence used to support the prosecution of sub-postmasters. Trust was also placed in the testimony of the Post Office and Fujitsu in maintaining that the Horizon system was reliable and not the source of the discrepancies. The consequences of their convictions were severe: financial ruin, criminal convictions and in some cases suicide. Yet in 2019, the truth of the matter, that the discrepancies in the accounts arise from bugs in the Horizon system was finally recognised. The cost of this mistrust in these cases is hard to summarise, but it is life and death for some. As Mason and Christie Blind say, trust in electronic evidence is a great weakness in the legal system for which the Law Commission, Judges and Lawyers must all work to address and accept some of the responsibility for this miscarriage of justice (MasonB, 2023, Christie 2020).

## 3.0 CONCLUSIONS AND AREAS FOR FUTURE RESEARCH

This paper aimed to explore a series of measures of trust that could be applied to generative AI and LLM produced spreadsheet formulas. This final section will consider the answers to the research questions posed and identify areas for future research.

### 3.1 Research Questions

**Research Question 1:** *What significant differences are there between trust dimensions for automatic machines and generative AI and LLMs?*

An analysis of different trust models provided by several key authors identifies a series of trust dimensions and evaluates the differences between generative AI, LLMs and automatic machines. These differences generally come in the freedom the user has to define the input to the generative AI process and the need to evaluate the output as opposed to automatic machines which generally have far fewer input variations, which are predefined, which are not defined by the operator, or have far fewer possibilities in the definition of the function of the machine. Other differences are apparent in the output of these different processes, generative AI and LLMs are prone to hallucinate and provide incorrect or nonsensical output.

**Research question 2**: *Can dimensions of trust in automation be adapted to evaluate spreadsheet formulas generated by AI and LLMs?*

The paper discusses a series of different dimensions and trust metrics that could help provide more objective evaluations of trust in AI generated spreadsheet formula through two groups of measures that examine the transparency and dependability of generative AI and LLM outputs.

Transparency is discussed in terms of the explainability of outputs generated and the visibility of the underlying processes that generate the output. Both aspects of transparency can be enhanced through engineering the prompt to provide robust reasoning and to provide details on the conditions under which the output was generated.

Dependability is discussed in terms of the reliability of the output and ethical considerations in the output. Reliability can be determined by validation and verification processes and through benchmarking different generative AI and LLM outputs with known tests as described in the literature. Ethical considerations deal with issues like bias, here use of external bias evaluation tools could help indicate where outputs display biases that generative AI and LLMs are known to display.

**Research Question 3**: *What sources of error and threats to trust exist in generative AI and LLMs and how might these issues be mitigated?*

Sources of error and threats to trust are explored in detail above, covering hallucinations, algorithmic bias and deficiencies in prompt engineering. Various possible solutions to these issues are explored including the use of prompt engineering libraries, rigorous validation and verification processes, the use of planning methods for prompt engineering and the implications of domain knowledge deficiencies in these activities.

### 3.2 Areas for future research

This paper has explored some of the dimensions of trust that could be applied to generative AI and LLM produced spreadsheet formula. There may be other dimensions of trust not discussed in this paper that could provide a useful evaluation of different dimensions of trust.

Planning methods such as data flow diagrams have significant potential in reducing omission errors in prompt engineering, but as yet little work exists in these areas. Agile methods like TDD could also have great utility in generative AI but similarly there is little research considering this. Both approaches could provide better prompts and may give a more reliable mechanism to generate spreadsheet formulas. A study examining the efficacy of these methods, and any other approaches to planning that might be useful, would provide an important part of the narrative.

The dimensions of trust offered could also be used to quantitatively calculate the risk or trustworthiness of any particular AI generated spreadsheet formulas. This process would consider the risks of generative AI such as hallucinations and balance the likelihood of that event happening against controls that can reduce the risk, for instance planning, the use of prompt libraries and verification. This could give a practical means to test the most critical uses of generative AI and ensure that basic steps to checking accuracy, bias and other issues.